\documentclass[conference]{IEEEtran}
\IEEEoverridecommandlockouts
\usepackage{cite,xurl}
\usepackage{amsmath,amssymb,amsfonts}
\usepackage{algorithmic}
\usepackage{graphicx}
\usepackage{textcomp}
\usepackage{xcolor}
\def\BibTeX{{\rm B\kern-.05em{\sc i\kern-.025em b}\kern-.08em
    T\kern-.1667em\lower.7ex\hbox{E}\kern-.125emX}}
\begin{document}

\title{Opportunities and Challenges of Urban Agetech:  from an Automated City to an Ageing-Friendly City 
}

\author{\IEEEauthorblockN{Seng W. Loke}
\IEEEauthorblockA{\textit{School of Information Technology} \\
\textit{Deakin University, VIC 3125, Australia} \\
seng.loke@deakin.edu.au}
} 

\maketitle

\begin{abstract}
Caring for the elderly, aging-in-place, and enabling the elderly to maintain a good life continue to be topics of increasing importance, especially in  countries with a higher percentage of older people, as people live longer, and care-giving costs rise.
This position paper proposes the concept of  {\em urban agetech}, where agetech services beyond the home can be an integral part of a modern ageing-friendly city, and where support for the elderly, where needed, in the form of automated systems (e.g., robots and automated vehicles) would be a normal city function/service, akin to the rather commonplace public transport services today.
\end{abstract}


\section{Two Visions of the Automated City}
What is an Automated City? One can think of increasing automation in public spaces in cities, from the emerging fully automated vehicles, urban robots and smart drones, to automated mass transport systems which are already in many cities throughout the world. The Automated City~\cite{quteprints213691} (and applications of AI to cities~\cite{aiandcity}) captures the idea of increasing automation in smart cities, somewhat paralleled by data-driven smart cities and participatory cities. 
One can imagine  city functions or services automated or partially automated, including  policing, waste collection and management, infrastructure maintenance, transportation, health and elder care, business transactions, citizen engagement, town services, and so on. 

Popular examples of automation include robots, of different forms and sizes.  If we take a broad view of {\em robots} as physically embodied Artificial Intelligence, one can ask two questions:

\begin{itemize}
    \item {\em How do we design a robot-friendly city?
(adding robots/machines to a human habitat) } This question might evoke science fiction scenes of humans living with robots in utopian and/or dystopian futures, but without going too far into the future, one can imagine automated cities with robots performing city functions and helping people go about their daily lives. While we will make machines that will fit into and adapt to human lives,  we might also design and plan cities that are robot-friendly in order that the benefits of robotics technology can be fully exploited wherever we live.

\item {\em How do we design a human-friendly robot-city? 
(adding humans to a ``city-sized'' machine)} What if we imagine a city as a large machine in which people live? That is, we use the metaphor of ``city as a machine''. Thinking of where we live as a machine is not  a new idea. Le Corbusier, a Swiss-French architect, designer, and urban planner,   in his 1923 book Vers Une Architecture ({\em Towards a New Architecture}), described a house, as ``a machine for living in.'' Of course, a conceptualization of a machine in 1923 would be very different from a conceptualization of a machine in 2024.  But the idea is that there are aspects of life which can be automated (e.g., sewerage disposal when we go to the toilet and what happens at the touch of a flush button, or turn the water tap on, or plug in a device), built into the very human habitat. Given more advanced technologies of today, we can think of the smart home, a machine in which we live in with many home functions automated. Scale this up to a smart building and then to smart collections of buildings linked by automated transportation systems, and we then live in a rather large ``machine'' comprising many parts (and the next step is to scale this to a city!). Bruce Schneier wrote that the ``Internet of Things Will Be the World's Biggest Robot''.\footnote{\url{https://www.schneier.com/blog/archives/2016/02/the_internet_of_1.html}}

\end{itemize}
From the two questions then come two visions of the automated city:
(a) a {\em robot-friendly} city, and (b) a {\em city as a machine for living in}. One might consider these as two sides of the same coin, where (a) focuses on how to add robotic services to an existing city, while (b) focuses on almost building an automated city (or part of a city) from ground-up if the possibility is there; if we think of a city-machine comprising multiple parts (e.g., a ``machine'' which is a smart building, or a smart system comprising a collection of buildings and connecting infrastructure, with built-in automation and sensing as well as hundreds of robots, coordinated, being components of the same building, or system, under centralised or decentralised but coordinated control).

\section{Agetech Today and the Future}
A related question is how such an automated city could help the elderly or the aged in their daily lives, especially those with some forms of ailments or (even partial) physical impairments, beyond providing simple conveniences and efficiencies. 

\begin{figure}[h]
\centering
\includegraphics[width=0.9\linewidth]{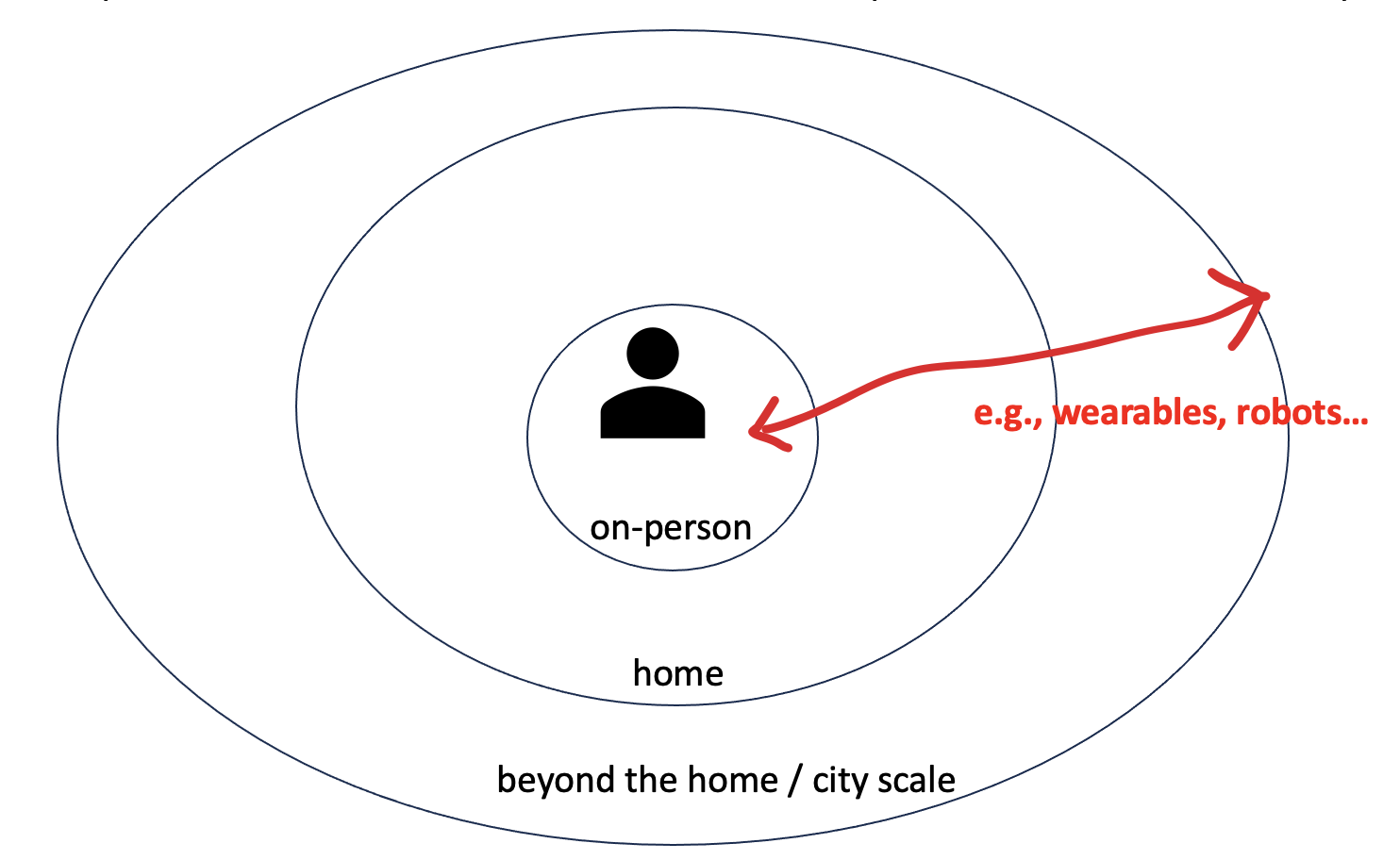}
\caption{Agetech beyond the home.}
\label{agetechbeyondhome}
\end{figure}

{\em Agetech} (a term often referring to digital technology for today's older adults; or  {\em gerontechnology}) today has enjoyed tremendous attention, growth and developments (e.g., see~\cite{etkin2022agetech}), and will continue to do so in many different categories~\cite{Liu22} including 
\begin{itemize}
    \item technologies to  support/enable autonomy and independence: preserving self-determination and free-will, and allowing one to make decisions or perform a task or an activity without help;
    \item technologies at home,  to support Activities of Daily Living (ADL), e.g., laundry, cooking, eating, cleaning, etc; and to 
support independent self-care and self-health management - e.g., health information management,  physiological and activity monitoring, exercise, and rehabilitation; and
    \item  technologies beyond the home, to support activities for economic and social participation, such as payment and access to finance, voting, interactions/friendships, remain connected,  
to support activities and hobbies, and so on, for enjoyment and self-fulfillment - e.g., physical activities, lifelong learning, spirituality, creative endeavours, gaming, communications, and to support independence in mobility / transport in the community -  e.g., location tracking, navigation, transportation to shops and so on.\footnote{For a review of related industries, see also the agetech market map: \url{https://thegerontechnologist.com/}}

\end{itemize}  
Hence, Agetech can comprise not just devices worn on the person or at home, but can cross boundaries of home and beyond the home, and cross boundaries of personal and social/community. Also, Agetech can be not just self-owned, but community-owned, or be ``part of'' the city in a way, that is, {\em urban agetech}; personal/self-owned devices can cooperate with community/city devices to help or support a person's activity beyond the home - see Figure~\ref{agetechbeyondhome}. For example, extrapolating from wheelchairs which could be borrowed in shopping malls, one can imagine concierge robots which can be borrowed to accompany an elderly person while shopping, perhaps even helping to carry items, and this combined with community automated vehicles might enable an elderly person to go from home to a shopping mall and back with robot help. Such a person could be helped by a bunch of technologies (effectively robots!) working together, each playing different roles. We can be thinking in terms of systems: a combination of technologies work together towards a task that an elderly person wants to accomplish.
We will explore such scenarios further in the next section.

\section{Scenarios and Personas}
We can consider the following task that an elderly person, called E, might want to do. Person E wants to go to the shops (several kms away) and shop,  and come home (and visit some friends along the way).
Person E is forgetful, slightly frail/not in good health, not rich, not tech-savvy and cannot use apps on a smartphone, generally alone (or children away), cannot drive and cannot carry too much weight.

An ideal is as follows: another person H takes person E to the shop, looks after E while E shops (even asks E: ``r u ok?'' from time to time) 
and helps E to rest when needed and helps E carry goods, and takes E home safely, and so on. But suppose there is no such H available (or costly to hire); so, how does E do it? 
Two possibilities with (emerging) technology are:
\begin{itemize}
   \item replace H by a robot R capable of all those things which H can do! (But we are not there yet?! We will return to this point later in discussing humanoid robots.)
\item replace H by a collection of robots and devices (each being able to do some task that H does; e.g., a health monitoring wearable device, 
an automated vehicle, a robot trolley to help carry goods, another device to help E navigate around etc), that is, we envision a collection of automated vehicles and robots that work together to help E complete the above task.  But how would E use all that?! Perhaps we can consider a technologically augmented built environment (physical infrastructure \& digital infrastructure) to be a {\em partner} or {\em host} to enable E to accomplish this task, which E is somehow able to interact with?
\end{itemize}

On this note, we can consider other scenarios and personas:
\begin{itemize}
\item Consider an 80+ year old elderly lady, very limited walking ability (100-200m), literate in English, Chinese and dialects, requires at least a walking stick, not tech-savvy (can use a smartphone to make calls and receive calls, with pre-programmed numbers, and not able to download and use any apps, etc – only basic phone features), unable to drive, can do payments with cash and withdraw cash from bank (unable to use credit card or any e-payment system), not able to cook anymore, medical condition requires monthly visits to doctor for checkup, and has limited strength to carry even groceries; the lady likes to go out to shopping malls and eat out, and have occasional visits from friends (who are all also around eighty years old) and children working and living overseas (she does not want to depend on her children); she is living alone, has a paid occasional domestic housekeeper, owns a house in a residential area 1.5km from the nearest shops (not quite reachable walking).
Question: how can she get around, shop, and live independently?

\item Consider a 70+ elderly female-male couple, both with limited walking ability (female can walk independently up to several hundred meters, male can walk a little further); both not tech-savvy, both literate in English and a dialect, both not able to cook anymore; male able to drive (though with eye-sight limitations, and slower reflex; can drive for short trips), and able to use cash (but no credit card) and smartphone (can use some apps such as Whatsapp and Facebook, but cannot use navigation apps such as Google Maps or on-demand transport apps such as Grab); male has limited strength to carry light groceries. The female can use phone and WhatsApp to make and receive calls but not any other apps, requires regular doctor checkups, and has mild dementia, limited strength and cannot carry groceries; they don't have  children  - the couple is living alone, has no domestic helper, owns a house in a new residential area 1-2km from the nearest shops (they now live different from where they use to live, are still not familiar with the area, and are, currently, too far from relatives and friends).
Question: how can they get around and shop around and live independently?
\item Consider a 75+ man, with physical impairments, generally needs wheelchair - not able to walk and unable to drive, but otherwise healthy, literate in Italian and knows some English, living alone in his own apartment, not tech-savvy but can use smartphone for phone calls; he has no children, and has a paid occasional domestic helper.
Question: how can he maintain social relationships and participate in the community?
\end{itemize}
The above are only three possible personas, and somewhat contrived, but one can think of many others. Each elderly person would be different, with varying preferences, resources and capacities, and there is no ``one-size-fits-all'' or ``one tech for all problems'', but the purpose of these personas above are to highlight possible opportunities for  automated city services to support, not just mobility around the home, but  support physical activities and performing of tasks in places beyond the home; of course, one could augment the above with help or support for ADLs required at home. Indeed,
one could think of automation and technological solutions to the above scenarios (and the reader is invited to brainstorm and think through robotic solutions for the above). And
there seems a challenge and opportunity for complementing at home automation with beyond home automation to provide a holistic support system.

So, the question then is whether automated help for the aged in such scenarios above could be designed and implemented as a ``given'' of a city, or a city function (or publically available service), in a way akin to the public bus service or train service which one might observe common in many cities. 

\section{Humanoid Robots as a ``Swiss Army Knife'' of Robots?}
Robots have been considered for elder care for some years now - e.g., see reviews~\cite{yuan23,zhang24,Wang2024}.\footnote{For an Australian viewpoint, see \url{https://www.ariia.org.au/sites/default/files/2023-02/ARIIA-technology-attitudes-towards-robots.pdf} and for a cautionary note of robot benefits: \url{https://www.nature.com/articles/d41586-024-01184-4} and \url{https://www.technologyreview.com/2023/01/09/1065135/japan-automating-eldercare-robots/}}
More recently,  there have emerged in popular media multiple examples of humanoid robots~\cite{JAS-2023-1103}, including Tesla's Optimus\footnote{\url{https://x.com/tesla_optimus}}, 1x Technologies' Neo\footnote{\url{https://www.1x.tech/androids/neo}}, Neurarobotics' 4NE-1\footnote{\url{https://neura-robotics.com/meet-4ne-1}}, Sanctuary's Phoenix\footnote{\url{https://sanctuary.ai/}}, Unitree Robotics' H1\footnote{\url{https://www.unitree.com/h1/}}, and Agility Robotics' Digit\footnote{\url{https://agilityrobotics.com/products/digit}}.
Can a humanoid robot play the role of H in the scenario of the previous section to help person E? This would include helping the aged person E from the door of the home into a (perhaps automated) vehicle and then taking E to the shops and then guiding E around, helping to carry shopping when required, and then accompanying E home. 

Humanoid robots are mentioned here as they hold promise for performing multiple functions well, to play different roles, in contrast to using a collection of separate robots, each for different purposes, i.e., metaphorically, a ``Swiss Army Knife'' robot. The use of such robots for elder care or assisting people with disabilities has been reviewed, e.g., in~\cite{doi:10.1080/10400435.2021.1880493,doi:10.1080/10400435.2024.2337194}.
The aid that a humanoid robot can provide might extend not only beyond the home but within the home. There have been videos, in popular media, of robot arms helping to cook burgers, make coffee and cleaning bathrooms; a humanoid robot with such arms might be able to do something similar. Such a robot (if human-sized) can even sit in a non-automated vehicle and drive it (at least as we see in some science fiction movies!), or accompany an elderly person on a bus.
 
 However, the development of humanoid robots is still on-going, but they can provide an {\em integrative platform} into which multiple types of AI and robotics technologies can be integrated (e.g., see~\cite{lcao2024ai}), including being instructable and conversational (in the chatGPT style), working as a co-bot for certain ADL tasks, within and beyond the home. The idea of a collection of humanoid robots for low ``rent'' (privately owned and/or government subsidied), or on a pay-as-you-use scheme, providing some of the support services above might seem far-fetched today, but perhaps not inconceivable. 
 
\section{Challenges and Conclusions}
This paper has reflected on how technology (including AI and IoT) can address issues of the aged in cities, proposing the notion of urban agetech, referring to publically available city-services or functions for the aged.
Such urban agetech can help extend the range of activities for the aged, facilitating even greater autonomy, increased independence, and greater accessibility to urban life. An elderly person might then be able to have walks, not just  around the home or around the immediate neighbourhood, but also  go shopping in other places kilometres away or even go on a tour in another country/unfamiliar city, with companion robots and automated vehicle for those who cannot drive anymore.

However, to accommodate such robots and automation in cities, city redesign might be considered (e.g., there has been much thought about the ramifications of automated vehicles for urban planning and design, e.g., \cite{doi:10.1080/01441647.2023.2189325}\footnote{See also \url{https://www.wired.com/2016/10/heres-self-driving-cars-will-transform-city/}}). Recent work has attempted to develop ISO standards for what has been called {\em Public-Area Mobile Robots} (PMRs),\footnote{For example, see \cite{Salvini20}, and \url{https://www.iso.org/standard/85782.html} and \url{https://www.urbanroboticsfoundation.org/}.}
which will suggest safety, societal, and ethical requirements on the behaviour of robots (e.g., delivery robots, cleaning robots, and so on) in public spaces.

\begin{figure*}[h]
\centering
\includegraphics[width=0.95\linewidth]{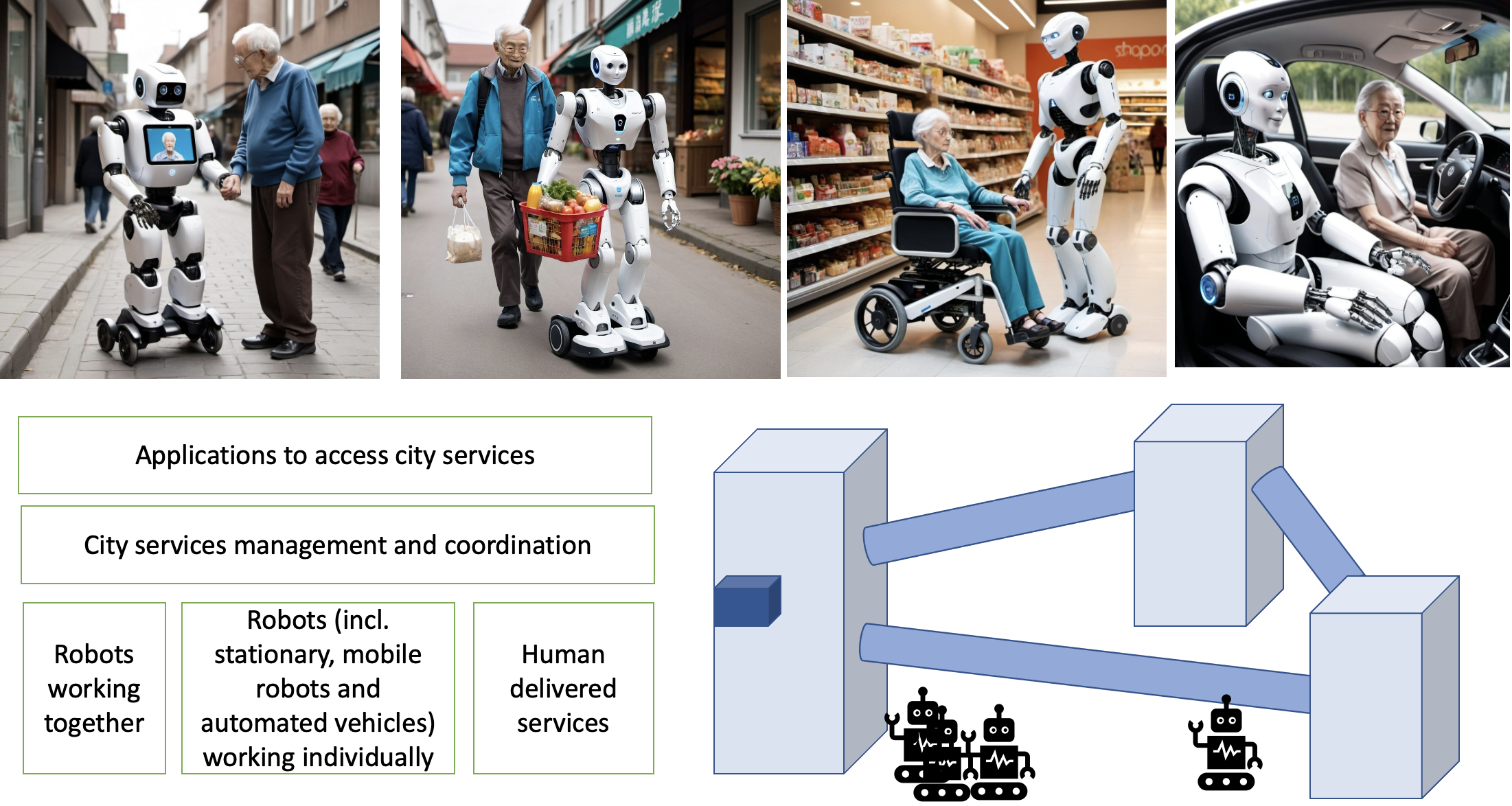}
\caption{The top row of images illustrates robots helping  the elderly beyond the home (the last image on the right shows a robot accompanying an elderly person who can still drive, but the robot can take over at any time) [images generated using https://openart.ai/]. The second row of diagrams first illustrates the layered architecture of apps (which need to be designed to be elderly friendly)  that allow access to city services/functions, and then the architectural layers of management and coordinated robot services to accomplish the city functions and services. The diagram on the right illustrates the vision of a city as an AI/machine for people to live in, with many components – a complex system of inter-woven cooperative robots/systems; effectively the automated aspects of a city is a robot with many Internet-connected disjointed/separate physical components but all connected to a central controller; scaling up a `machine for living in': living in a smart home, a smart building, and a collection of smart buildings with connecting autonomous transport infrastructure, and eventually, a smart (automated) city.}
\label{vision}
\end{figure*}

  Figure~\ref{vision} illustrates the vision of urban agetech beyond the home, supported by robots (and humans) working together and/or individually, and  architectural layers required. The apps must be designed to be elderly friendly and accessible to people with partial impairments.

 There are a number of considerations to introducing automated or robotic services as public services:
\begin{itemize}
    \item {\em human-centric}: (a) supporting people in their daily life activities and work with humans~\cite{vianello:hal-03413650,strutz24}, (b) fairness of access to the services, including wait-times and being economical, (c) ethical and conforming to societal norms, (d) personalized support, and (e) humans maintaining (some) control of services, usable by the elderly and yet humans remain the loop for management;
    \item {\em resilient to failures}: local failure should only have local impacts, and there needs to be alternatives and quick recovery in case services break down and critical services need to have high reliability;
        \item {\em easy maintenance}: the ``infrastructure'' of automated services must be maintenable and upgradable;
    \item {\em contextualized}: each city and area within the city has its own requirements and characteristics, and must be usable by the aged (from UI design to designing for usability) perhaps demographically influenced; 
    \item {\em evolvable and compositional}: the services should  be incrementally extendable,  and can be built up over time via adding components (e.g., adding robots where the newer robots should be able to work seamlessly with the older ones, or new sensors can be utilized by existing robots to improve what they normally already do); perhaps useful for this is the idea of robots being able to cooperate with each other (e.g., via robot-to-robot communications) and the need for standards in protocols for cooperation (e.g., when robots belong to different manufacturers and owners,  occupy shared public spaces, and are programmed with different owner-interests) and societal expectations;
    \item {\em city-scale coordination}: coordination of robots in public spaces at city-scale might be required (e.g., including rules for robots in public~\cite{DBLP:journals/ais/Loke24}) if such services are to be commonplace, requiring multiparty considerations, including industry, government and technologists; and 
    \item {\em aesthetics}: robots and automated vehicles (even drones) in cities should not take away aesthetics from the city, but could contribute to it, similar to the role of architecture for our built environments.
\end{itemize}

While we have highlighted humanoid robots as an integrative platform to provide such urban agetech services, there could be different realizations for urban agetech services, from collections of different types of specialised robots to different human-robot collaborative arrangements, perhaps more practical in the days when humanoid robots  and AI are still developing.

While urban agetech cannot replace the human touch and the human element, and there is the need for human acceptance of the technology, the hope is that some of the challenges of growing old can be addressed with technology, so that humans can focus better on what only they can do best.

\section*{Acknowledgment}
The author would like to thank Professor Belinda Yuen and colleagues from the Lee Kuan Yew Centre for Innovative Cities (LKYCIC) at the Singapore University of Technology and Design, for conversations where most of the ideas from this paper were first presented.


\bibliographystyle{IEEEtran}
\bibliography{refs}

\end{document}